\documentclass[useAMS,usenatbib]{mn2e}
\usepackage{url,times,graphicx,amsmath,amsfonts,amssymb,aas_macros,color,epsfig,epstopdf}
 \usepackage{xcolor}

\newcommand{\hMpc}{{\ifmmode{h^{-1}{\rm Mpc}}\else{$h^{-1}$Mpc}\fi}}
\newcommand{\hkpc}{{\ifmmode{h^{-1}{\rm kpc}}\else{$h^{-1}$kpc}\fi}}
\newcommand{\kpc}{{\ifmmode{ {\rm kpc} }\else{{\rm kpc}}\fi}}
\newcommand{\kms}{{\ifmmode{ {\rm km\,s^{-1}} }\else{ ${\rm km\,s^{-1}}$ }\fi}}
\newcommand{\hMsun}{{\ifmmode{h^{-1}{\rm {M_{\odot}}}}\else{$h^{-1}{\rm{M_{\odot}}}$}\fi}}
\newcommand{\Msun}{{\ifmmode{{\rm M}_{\odot}}\else{${\rm M}_{\odot}$}\fi}}
\newcommand{\Mhalo}{{\ifmmode{M_{\rm halo}}\else{$M_{\rm halo}$}\fi}}
\newcommand{\Rvir}{{\ifmmode{R_{\rm vir}}\else{$R_{\rm vir}$}\fi}}
\newcommand{\Mstar}{{\ifmmode{M_{\star}}\else{$M_{\star}$}\fi}}
\newcommand{\Vrot}{{\ifmmode{V_{\rm rot}}\else{$V_{\rm rot}$}\fi}}
\newcommand{\ltsima}{$\; \buildrel < \over \sim \;$}
\newcommand{\gtsima}{$\; \buildrel > \over \sim \;$}
\newcommand{\lsim}{\lower.5ex\hbox{\ltsima}}
\newcommand{\gsim}{\lower.5ex\hbox{\gtsima}}

\def\lesssim{\mathrel{\hbox{\rlap{\hbox{\lower4pt\hbox{$\sim$}}}\hbox{$<$}}}}
\def\gtrsim{\mathrel{\hbox{\rlap{\hbox{\lower4pt\hbox{$\sim$}}}\hbox{$>$}}}}

\newcommand{\Sec}[1]{Section~\ref{#1}}
\newcommand{\Eq}[1]{Eq.~(\ref{#1})}
\newcommand{\Fig}[1]{Fig.~\ref{#1}}
\newcommand{\beq}{\begin{equation}}
\newcommand{\eeq}{\end{equation}}
\def\beqa{\begin{eqnarray}}
\def\eeqa{\end{eqnarray}}

\def\head{ \vbox to 0pt{\vss \hbox to 0pt{\hskip 440pt\rm
      LA-UR-10-07069\hss} \vskip 25pt}}

\def\head{
 \vbox to 0pt{\vss
                   \hbox to 0pt{\hskip 440pt\rm LA-UR-10-07069\hss}
                  \vskip 25pt}}
\title[Cusps and Cores in MaGICC galaxies]
{The dependence of dark matter profiles on the stellar to halo mass ratio: a prediction for cusps vs cores}

\author[Di Cintio et. al]
       {Arianna Di Cintio$^{1,2}$\thanks{E-mail: arianna.dicintio@uam.es}, Chris B. Brook$^1$, Andrea V. Macci\`{o}$^3$, Greg S. Stinson$^3$, \newauthor Alexander Knebe$^1$, Aaron A. Dutton$^{3}$, James Wadsley$^4$ \\
$^{1}$Departamento de F\'isica Te\'orica, M\'odulo C-15, Facultad de Ciencias, Universidad Aut\'onoma de Madrid, 28049 Cantoblanco, Madrid, Spain\\
$^2$Physics Department G. Marconi, Universit\`{a} di Roma Sapienza, Ple Aldo Moro 2, 00185 Rome, Italy\\
$^3$Max-Planck-Institut f\"ur Astronomie, K\"onigstuhl 17, 69117 Heidelberg, Germany\\
$^4$McMaster University, Hamilton, Ontario, L8S 4M1, Canada\\
}

\begin{document}

\date{Accepted 2013 October 03. Received 2013 September 16; in original form 2013 June 04}

\pagerange{\pageref{firstpage}--\pageref{lastpage}} \pubyear{2010}

\maketitle

\label{firstpage}


\begin{abstract}

  We use a suite of 31 simulated galaxies drawn from the MaGICC
  project to investigate the effects of baryonic feedback on the density profiles of dark matter haloes. The sample
  covers a wide mass range: $9.4\times 10^{9} < \Mhalo / \Msun < 7.8\times 10^{11}$, hosting galaxies with stellar masses:
  $5.0\times 10^{5} < \Mstar/\Msun < 8.3\times 10^{10}$, i.e. from
  dwarf to L$^\star$. The galaxies are simulated with blastwave supernova feedback and, for some of them, an additional source of energy from massive stars is included.
 Within this feedback scheme we vary several parameters, such as the initial mass function, the density threshold for star formation and energy from supernovae and massive stars.

     The main result is a clear dependence of the inner slope of
  the dark matter density profile, $\alpha$ in $\rho$\,$\propto$\,$r^\alpha$, on
  the ratio between stellar-to-halo mass, $\Mstar / \Mhalo$. This relation is independent of the particular choice of parameters within our stellar feedback scheme, allowing a prediction
  for cusp vs core formation.
  When $\Mstar / \Mhalo$ is low, $\lsim  0.01$ per cent, energy from stellar feedback is insufficient to significantly alter the inner dark matter density and the galaxy retains a
  cuspy profile. At higher ratios of stellar-to-halo
  mass feedback drives the expansion of the dark matter and generates cored
  profiles. The flattest profiles form where
  $\Mstar/\Mhalo \sim 0.5$ per cent. Above this ratio, stars
  formed in the central regions deepen the gravitational potential 
  enough to oppose the supernova-driven expansion
  process, resulting in cuspier profiles.
  Combining the dependence of $\alpha$ on $\Mstar / \Mhalo$ with
  the empirical abundance matching relation
  between $\Mstar$ and $\Mhalo$ provides a prediction for how 
  $\alpha$ varies as a function of stellar mass. Further, using the Tully-Fisher 
  relation allows a prediction for the dependence of the dark matter inner slope on the observed rotation velocity of
  galaxies. The most cored galaxies are expected to have
  $\Vrot\sim 50\, \kms$, with $\alpha$ decreasing for
  more massive disc galaxies: spirals with
  $\Vrot\sim 150\,\kms$ have central slopes $\alpha\leqslant-0.8$, approaching again the NFW profile.
 This novel prediction for the dependence of $\alpha$ on disc galaxy mass can be tested using observational data sets and can be applied to theoretical modeling of mass profiles and populations of disc galaxies.\end{abstract}

\noindent
\begin{keywords}

cosmology: dark matter galaxies: evolution - formation - hydrodynamics methods:N-body simulation 

 \end{keywords}

\section{Introduction} \label{sec:introduction}
The $\Lambda$ Cold Dark Matter ($\Lambda$CDM) cosmological model has been shown to agree with observations of structures on large scales \citep[e.g.][]{Riess98,Komatsu11,Hinshaw12,Planck13}.
According to this theory, galaxies are embedded within dark matter (DM) haloes \citep{whiterees78,blumenthal84}, whose properties have been extensively studied in the past thanks to numerical N-body simulations \citep[e.g.][]{Springel05,Power06,maccio08,Kuhlen12}.
Problems at small scales, however, still affect the $\Lambda$CDM model, one of which is the so-called ``cusp-core" problem. A prediction of pure DM collisionless simulations is that dark matter density increases as $\rho$\,$\propto$\,$r^{-1}$ toward the halo center \citep{Navarro96,Springel08,Navarro10}. The existence of such a ``cuspy" density profile is in disagreement with observations of disc and dwarf galaxies
\citep[e.g.,][]{Salucci00,Simon05,Deblok08,Kuzio08,Kuzio09,oh11b}, where detailed mass modeling using rotation curves suggests a flatter, or ``cored", DM density profile. 
Simulated DM haloes modeled with an Einasto \citep{Einasto65} profile have a inner slope of $-0.7$ \citep{Graham06}: this value is closer to what observed in real galaxies \citep{Swaters03}, yet not sufficient to solve the discrepancy \citep{deBlok03}.

One possibility, without resorting to more exotic forms of dark matter (e.g. warm dark matter see \citealt[][\citeauthor{Maccio2012b} 2012b]{AvilaReese01,Bode01,Knebe02}), is that this inconsistency arises from having neglected the effects of baryons, which are irrelevant on cosmological scales where dark matter and dark energy dominate, but may be dynamically relevant on small, galactic scales. For example, as gas cools to the central region of galaxy haloes, it adiabatically contracts dark matter to the centre \citep[e.g.][]{Blumenthal86,Gnedin04}. Such adiabatic contraction exacerbates the mismatch between the profiles of dark matter haloes and the observed density profiles inferred from rotation curves. Further, theoretical models with halo contraction are unable to
self-consistently reconcile the observed galaxy scaling relations, such as
the rotation velocity-luminosity and size-luminosity relations.
Un-contracted or expanded haloes are required \citep{Dutton07,Dutton2011}. 

Two main mechanisms have been shown to cause expansion: supernova feedback  \citep{Navarro96b,Mo04,Read05,Mashchenko06,Pontzen12} and dynamical friction \citep{El-Zant01,Tonini06,Romano-Diaz08,Goerdt10,Cole11}. Supernova feedback drives sufficient gas outflows to flatten the central dark matter density profile in simulated dwarf galaxies \citep{governato10,teyssier13} into a ``core''.  
Dynamical friction smooths dark matter density profiles during mergers.

The analytical model of \citet{Pontzen12} predicts that repeated outflows, rather than a single, impulsive mass loss (as in \citealt{Navarro96b}), transfer energy to the dark matter.  The rapid oscillations of the central gravitational potential perturb the dark matter orbits, creating a core. \citet{Mashchenko06} decribed a similar mechanism in which supernova-driven outflows changed the position of the halo centre, also creating a core. \citet{Maccio12} showed that reasonable amounts of feedback in fully cosmological simulations can result in dark matter cores rather than cusps in galaxies as massive as L$^\star$.
\cite{Governato12} measured the inner dark matter slope in a sample of simulated dwarf galaxies, that match well the stellar-to-halo mass relation \citep{Munshi13}, using a power law density profile $\rho$\,$\propto$\,$r^\alpha$.  They found that the slope $\alpha$ increases, i.e. the profile flattens, with increasing stellar mass.

In this paper we study dark matter density profiles in a suite of galaxies drawn from the MUGS \citep{stinson10} and MaGICC projects \citep{stinson13,brook12b}. The galaxies cover a broad mass range from dwarf to massive discs, and are simulated using a variety of stellar feedback implementations. 
The wide mass range of our simulated galaxies, $5.0 \times 10^{5}<\Mstar/\Msun < 8.3 \times 10^{10}$, allows us to confirm and extend the results of \citet{Governato12}. We show that the most relevant property for the determination of the DM inner slope is actually the stellar-to-halo mass ratio, i.e. the star formation efficiency, and that the relation between $\alpha$ and stellar mass turns over such that the inner density profiles of more massive disc galaxies become increasingly steep. 

We present our simulations in \Sec{sec:simulation}, the results and predictions in \Sec{sec:results} and the conclusions in \Sec{sec:conclusion}.

\section{Simulations} \label{sec:simulation}


\begin{table*}
\begin{minipage}{180mm}
\begin{center}
\caption{Simulation parameters}
\begin{tabular}{lllllllllllll}
\hline

&MUGS  &gas part. & soft &$\Mhalo$ &$\Rvir$ &$\Mstar$ &  $E_{\rm SN}$& $\epsilon_{\rm esf}$&IMF & $n_{\rm th}$ &sym \\
&      label &mass [M$_\odot$]  & [pc]   &   [M$_\odot$]& [kpc]& [M$_\odot$]& & & & [cm$^{-3}$]  & & \\
\hline

     Low & g1536      &   3.1$\times$10$^3$& 78.1  		&$9.4$$\times$10$^{9}$& 61& $7.2$$\times$10$^{5}$& 1.0 &0.1 &C & 9.3&   $ \color{red} \bullet $  \\
& g1536      &   3.1$\times$10$^3$& 78.1 	&$9.4$$\times$10$^{9}$& 60& $5.1$$\times$10$^{5}$&1.0 &0.125 &C &9.3 &$\color{blue}  \bullet$ \\
& g1536      &   3.1$\times$10$^3$&  78.1	&$9.4$$\times$10$^{9}$& 61& $5.0$$\times$10$^{5}$& 1.0 &0.175 &C &9.3 &$\color{yellow}  \bullet $\\
& g1536      &   3.1$\times$10$^3$& 78.1 	&$9.4$$\times$10$^{9}$& 60& $7.0$$\times$10$^{5}$& 1.2& 0.0&C &9.3 &$\blendcolors{!50!white}\color{cyan}   \bullet $ \\
 &g15784 &  3.1$\times$10$^3$   &78.1 		& $1.9$$\times$10$^{10}$ & 77&  $8.9$$\times$10$^{6}$&  1.0 &0.1 &C & 9.3&   $ \color{red} \blacktriangle  $ \\
  &g15784 &  3.1$\times$10$^3$   & 78.1 		& $1.9$$\times$10$^{10}$ &79 &  $7.4$$\times$10$^{8}$&  0.4 &0  &K &0.1 & $ \blacktriangle $\\
    &g15784 &  3.1$\times$10$^3$   & 78.1 		&  $1.9$$\times$10$^{10}$ & 79&  $8.4$$\times$10$^{6}$&  1.0 &0.125 &C &9.3 &$\color{blue}  \blacktriangle$\\
    &g15784 &  3.1$\times$10$^3$   & 78.1 	& $1.8$$\times$10$^{10}$ & 75&  $6.0$$\times$10$^{6}$&  1.0 &0.175 &C &9.3 &$\color{yellow}  \blacktriangle $\\
     &g15784 &  3.1$\times$10$^3$   &  78.1	& $1.8$$\times$10$^{10}$ & 75&  $1.1$$\times$10$^{7}$& 1.2& 0.0&C &9.3 &$\blendcolors{!50!white}\color{cyan}   \blacktriangle$\\
 &g15807 &  3.1$\times$10$^3$  &78.1 		&$3.0$$\times$10$^{10}$ & 89& $1.6$$\times$10$^{7}$ &  1.0 &0.1 &C & 9.3&   $ \color{red} \blacksquare $\\
\hline
 Medium &g7124    &  2.5$\times$10$^4$    & 156.2 &  $5.3$$\times$10$^{10}$ & 107  &$1.3$$\times$10$^{8}$ &  1.0 &0.1 &C & 9.3&   $ \color{red} \ast $\\
 &g5664    &  2.5$\times$10$^4$   &  156.2	&$6.3$$\times$10$^{10}$ & 114 & $2.4$$\times$10$^{8}$ & 1.0&0.1 &C &9.3 &$ \color{red}  \blacklozenge $ \\
   &g5664    &  2.5$\times$10$^4$   & 156.2 	& $6.6$$\times$10$^{10}$ & 116 & $1.0$$\times$10$^{9}$ & 0.8 &0.05  &C &9.3 & $ \color{violet} \blacklozenge $\\

 &g5664    &  2.5$\times$10$^4$   & 156.2 &  $7.3$$\times$10$^{10}$ & 120& $8.7$$\times$10$^{9}$ & 0.4 &0  &K &0.1 & $ \blacklozenge $\\

 &g1536    &  2.5$\times$10$^4$    & 156.2 	&$8.3$$\times$10$^{10}$ &  125 &$4.5$$\times$10$^{8}$ &  1.0 &0.1 &C & 9.3&   $ \color{red} \bullet $  \\
 &g21647    &  2.5$\times$10$^4$    &  156.2&  $9.6$$\times$10$^{10}$ & 131  &$2.0$$\times$10$^{8}$ &  1.0 &0.1 &C & 9.3&   $ \color{red} \blacktriangleleft  $ \\
 &g15784 &  2.5$\times$10$^4$   &   156.2	&$1.8$$\times$10$^{11}$ & 161&  $4.3$$\times$10$^{9}$& 1.0 &0.1 &C & 9.3&   $ \color{red} \blacktriangle  $   \\
  &g15784 &  2.5$\times$10$^4$   & 156.2  	& $1.8$$\times$10$^{11}$ & 161&  $2.4$$\times$10$^{9}$& 1.0 &0.125 &C &9.3 &$\color{blue}  \blacktriangle$ \\
  &g15784 &  2.5$\times$10$^4$   & 156.2  	&$1.9$$\times$10$^{11}$ & 164 &  $7.1$$\times$10$^{9}$& 1.0 &0.1 &K &9.3 &$\color{orange}  \blacktriangle$ \\
  &g15784 &  2.5$\times$10$^4$   & 156.2  	& $1.7$$\times$10$^{11}$ &157 &  $8.6$$\times$10$^{8}$& 1.0 &0.1 &C &9.3 &$\color{green}  \blacktriangle$ \\
 &g15807 &  2.5$\times$10$^4$  & 156.2 	&$2.9$$\times$10$^{11}$ & 189 & $1.5$$\times$10$^{10}$ & 1.0 &0.1 &C & 9.3&   $ \color{red} \blacksquare $ \\
  \hline
  
   High &g7124    &  2$\times$10$^5$    & 312.5 &     $4.5$$\times$10$^{11}$ & 219 &$6.3$$\times$10$^{9}$ & 1.0 &0.1 &C & 9.3&   $ \color{red} \ast $\\
 &g7124    &  2$\times$10$^5$    & 312.5    &       $4.9$$\times$10$^{11}$ & 227 &$5.1$$\times$10$^{10}$ & 0.4 & 0&K & 0.1&  $ \ast $ \\
 & g5664    &  2$\times$10$^5$   & 312.5    &         $5.6$$\times$10$^{11}$ & 236 & $2.7$$\times$10$^{10}$ & 1.0&0.1 &C &9.3 &$ \color{red}  \blacklozenge $ \\
 & g5664    &  2$\times$10$^5$   &  312.5   &          $5.7$$\times$10$^{11}$ & 237 & $4.9$$\times$10$^{10}$ & 0.4 &0  &K &0.1 & $ \blacklozenge $\\
 & g5664    &  2$\times$10$^5$   & 312.5         &         $5.9$$\times$10$^{11}$ &241 & $1.4$$\times$10$^{10}$ &1.0 &0.175 &C &9.3 &$\color{yellow}  \blacklozenge $ \\
 &g1536    &  2$\times$10$^5$    &  312.5   &$7.2$$\times$10$^{11}$ & 257 &$2.4$$\times$10$^{10}$ & 1.0 &0.1 &C & 9.3&   $ \color{red} \bullet $  \\
  &g1536    &  2$\times$10$^5$    &  312.5&  $7.7$$\times$10$^{11}$ & 264 &$8.3$$\times$10$^{10}$ & 0.4 & 0&K & 0.1&  $ \bullet $ \\
   &g1536    &  2$\times$10$^5$    &  312.5& $7.0$$\times$10$^{11}$ & 254 &$1.1$$\times$10$^{10}$& 1.0 &0.125 &C &9.3 &$\color{blue}  \bullet$\\
   &g1536    &  2$\times$10$^5$    & 312.5 	&  $7.8$$\times$10$^{11}$ & 265 &$2.5$$\times$10$^{10}$ &1.0 &0.175 &C &9.3 &$\color{yellow}  \bullet $  \\
     &g1536    &  2$\times$10$^5$    &  312.5	& $7.0$$\times$10$^{11}$ & 255 &$1.8$$\times$10$^{10}$ & 1.2& 0.0&C &9.3 &$\blendcolors{!50!white}\color{cyan}  \bullet $  \\
   
   \hline  
\end{tabular}\\
\end{center}
\label{tab:data}
\end{minipage}
\end{table*}

The simulations used in this study are taken from the McMaster Unbiased Galaxy Simulations (MUGS: 
\citealt{stinson10}), which is a sample of 16 zoomed-in regions where $\sim$L$^\star$ galaxies form in a cosmological volume 68 Mpc on a side.  MUGS uses a $\Lambda$CDM cosmology with $H_0$= 73 \kms Mpc$^{-1}$, $\Omega_{\rm{m}}=0.24$, $\Omega_{\Lambda}=0.76$, $\Omega_{\rm{bary}}=0.04$ and $\sigma_8 = 0.76$ \citep[WMAP3,][]{Spergel07}.

All of the simulations are listed in Table 1 where they are separated into 3 mass groups: high, medium and low mass. The symbol shapes denote simulations with the same initial conditions, while the colors indicate the specific star formation and feedback model used. The medium and low mass initial 
conditions are scaled down variants of the high mass initial conditions, so that rather than residing in a 68\,Mpc cube, they lie 
within a cube with 34\,Mpc sides (medium) or 17\,Mpc sides (low mass). This rescaling allows us to compare galaxies with exactly the same merger histories at three different masses.
Differences in the underlying power spectrum that result from this 
rescaling are minor \citep{Springel08,maccio08,kannan12}. Moreover, as shown through the paper, this methodology does not affect our analysis and results since we reach, at the low halo mass end where we have made the rescaling, the same conclusions as in \citet{Governato12} whose galaxies do not have rescaled initial conditions.

Our galaxies were simulated using \textsc{Gasoline} 
\citep{wadsley04}, a fully parallel, gravitational N-body + smoothed 
particle hydrodynamics (SPH) code.  Cooling via hydrogen, helium, and 
various metal-lines in a uniform ultraviolet ionising background is included as described in \citet{shen10}. 

In addition to the hydrodynamic simulations, collisionless, dark matter-only simulations were performed for each initial condition. These DM-only runs exhibit a wide range of concentrations, from those typical of the L$^\star$ to dwarf galaxies. The concentration, $c$, varies between $10\lesssim c\lesssim 15$, where $c\equiv R_{\rm{vir}}/r_{\rm{s}}$ and $r_{\rm{s}}$ is the scale radius of the NFW profile \citep{Navarro96}. Such a range is sufficient to study density profiles. Indeed, the sample includes a number of galaxies with high $c$ at each mass range, a legacy of preferentially simulating galaxies with early formation times in order to model Milky Way formation. 
 
The main haloes in our simulations were identified using the MPI+OpenMP hybrid halo finder \texttt{AHF}\footnote{http://popia.ft.uam.es/AMIGA} \citep{Knollmann09,Gill04a}. \texttt{AHF} locates local over-densities in an adaptively smoothed density field as prospective halo centers. For a discussion of its performance with respects to simulations including baryonic physics we refer the reader to \citet{Knebe13}.
The virial masses of the haloes, \Mhalo, are defined as the masses within a sphere containing $\Delta=390$ times the cosmic background matter density at $z=0$. 

\subsection{Star Formation and Feedback}
The hydrodynamic simulations all include star formation, with the stars feeding energy back into the interstellar medium (ISM) gas. A range of star formation and feedback parameters are used in this
study: all of them employ blastwave
supernova feedback \citep{stinson06}, and some also include ``early stellar feedback'', the energy that massive stars release prior to their explosions as supernovae
\citep{stinson13}.

In all simulations, gas is eligible to form stars when it reaches
temperatures below 15000 K in a dense
environment, $n > n_{\rm th}$. Two different density thresholds are
used for star formation, $n_{\rm th}$=0.1 and 9.3 cm$^{-3}$. Gas denser than $n_{th}$ is converted to stars according to the \citet{kennicutt98} 
Schmidt Law:
\begin{equation}
\frac{\Delta M_\star}{\Delta t} = c_\star \frac{m_{\rm gas}}{t_{\rm dyn}}
\end{equation}
where $\Delta M_\star$ is the mass of the stars formed in $\Delta t$,
the time between star formation events (0.8\,Myr in these
simulations), $m_{\rm gas}$ is the mass of the gas particle, $t_{\rm
  dyn}$ is the gas particle's dynamical time, and $c_\star$ is the
fraction of gas that will be converted into stars during $t_{\rm dyn}$.

Supernova feedback is implemented 
using the \citet{stinson06} blastwave formalism, depositing 
E$_{\rm{SN}}\times$$10^{51}$~erg into the surrounding ISM at the end of the 
lifetime of stars more massive than 8\,M$_\odot$.  
Since stars form from dense gas, this energy would be quickly radiated away due to the 
efficient cooling. For this reason, cooling is disabled for particles 
inside the blast region.
Metals are ejected from Type~II supernovae (SNeII), Type~Ia
supernovae (SNeIa), and the stellar winds driven from asymptotic giant
branch (AGB) stars, and distributed to the nearest gas particles using
the smoothing kernel \citep{stinson06}. The metals can diffuse between gas particles as described in \citep{shen10}.

Early stellar feedback is included in most of our 
simulations.  It uses a fraction, $\epsilon_{\rm esf}$, of the total luminosity emitted by massive stars.  The luminosity of stars is modelled with a simple fit of the 
mass-luminosity relationship observed in binary systems 
\citep{torres10}:
\begin{equation}
\frac{L}{L_\odot} = 
\begin{cases}
\phantom{100\,}(M/\Msun)^{4},  & M < 10\,\Msun \\
100\, (M/\Msun)^{2},   & M > 10\,\Msun \\
\end{cases}
\end{equation}
Typically, this model corresponds to the emission of $2\times10^{50}$~erg per \Msun\ of the entire
stellar population over the $\sim$4.5\,Myr between a star's formation
and the commencement of SNeII in the region. These photons do not
couple efficiently with the surrounding ISM \citep{freyer06}.  To
mimic this highly inefficient energy coupling, we inject
$\epsilon_{\rm esf}$ of the energy as thermal energy in the surrounding
gas, and cooling is \emph{not} turned off. Such thermal energy
injection is highly inefficient at the spatial and temporal resolution
of cosmological simulations \citep{katz92,kay02}, as the
characteristic cooling timescales in the star forming regions are
lower than the dynamical time. In the fiducial model used in the MaGICC
simulations, $\epsilon_{\rm esf}$=0.1, which corresponds to the fraction
of ionizing UV flux emitted from young stellar populations.

Two initial mass functions were used in the simulations. MUGS used \cite[][denoted K]{kroupa93}, while most of the rest used
\cite[][denoted C]{chabrier03}.  \cite{chabrier03} produces two times more
type SNII per mass of stars born. 

The fiducial feedback (red colored symbols)
includes early stellar feedback with $\epsilon_{\rm esf}=0.1$, $10^{51}$erg of energy deposited per
supernova and a \citet{chabrier03} IMF. 
The early stellar feedback efficiency $\epsilon_{\rm esf}$ is increased from 0.1 to 0.125 (blue) in some simulations, while in others $\epsilon_{\rm esf}=0$, but the energy per supernova is then increased by 20 per cent (cyan). In yellow, we include simulations with $\epsilon_{\rm esf}$ = 0.175, in which diffusion of thermal energy from gas particles \citep{Stinson12,Wadsley08} is allowed to occur during the adiabatic expansion phase. We also include simulations made with the original MUGS feedback, with $4\times10^{50}$erg per supernova, a \citet{kroupa93} IMF and no $\epsilon_{\rm esf}$, which systematically overproduce the number of stars at each halo mass (black). 
Finally, an intermediate feedback implementation with $\epsilon_{\rm esf}=0.05$, Chabrier IMF and $8\times10^{50}$erg per supernova, has been also added (purple).

The reader is referred to \citet{stinson13} for a study of the effects of the parameters on the galaxy properties. Suffice to say that the fiducial simulations best match present observed galaxy properties \cite[see also][]{brook12b}.

\section{Results}\label{sec:results}

\begin{figure}
  \includegraphics[width=3.0in]{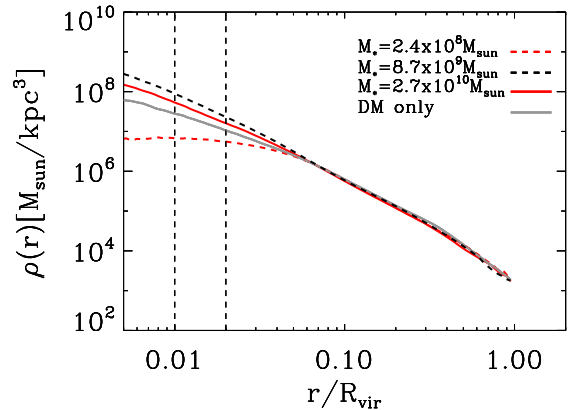}
  \caption{Density profiles of contracted (solid red and dashed black lines) and expanded (dashed red line) dark matter haloes, together with the corresponding DM only prediction (solid grey).  The vertical dashed lines indicate $0.01$ and $0.02$ of the virial radius, our fiducial range to measure $\alpha$. }
\label{fig:profiles}
\end{figure}

We study the response of the dark matter distribution to different
feedback schemes within this full set of simulated galaxies.  
Some example density profiles are shown in Figure \ref{fig:profiles}.
It shows how the dark matter density profiles of the hydrodynamic simulations 
can vary depending on physics (MUGS in black compared to MaGICC fiducial simulations, that use early stellar feedback, in red), galaxy mass (solid line at high mass and 
dashed line at medium mass), and how the hydrodynamic simulations compare with the
dark matter only run (solid grey line). 

The halo profiles are calculated using logarithmically spaced bins
and the dark matter central density is subsequently fit using a single power law, $\rho$\,$\propto$\,$r^\alpha$, over a limited radial range. 
The vertical dashed lines in \Fig{fig:profiles} show the fiducial range over which $\alpha$ is measured, $0.01<r/\Rvir<0.02$, where $\Rvir$ is the virial radius. Other radial ranges are also used to ensure the robustness of our results. 

The choice of $0.01\Rvir$ as the inner most bin satisfies the \citet{Power03} criterion for 
convergence even in our least resolved galaxy, as it encloses enough DM particles to ensure that 
the collisional relaxation time is longer than the Hubble time. 
This range is also straightforward to reproduce, and is not dependent on the
resolution of the simulations.
We also measured $\alpha$ between $3<r/\epsilon<10$, where $\epsilon$ is the softening length of each galaxy, and at a fixed
physical range, $1< r/\kpc<2$. The choice of radial fitting range does not affect
our results qualitatively, and only makes small quantitative
differences which we show in our main results.

\subsection{Inner slope as a funcion of halo mass}

\begin{figure}
  \includegraphics[width=3.4in]{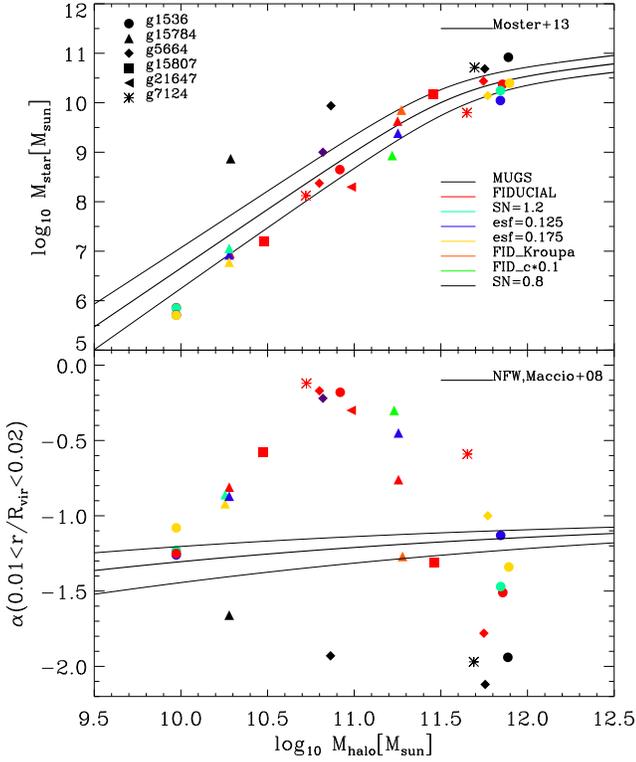}
  \caption{Top panel: The abundance matching relation for our suite of simulated galaxies. The feedback schemes are indicated with different colors, while the different galaxies are represented with symbols. The thick solid line corresponds to the abundance matching prediction from \citet{Moster13} and the thin lines are the $1\sigma$ uncertainty on it. Bottom panel: The inner slope of the dark matter distribution, measured between $0.01$ and $0.02$ of each galaxy's virial radius, as a function of total halo mass. The solid lines are the theoretical expectation for dark matter haloes from \citet{maccio08} with its scatter.}
\label{fig:AM_moster}
\end{figure}

We first examine how $\alpha$ varies with stellar and halo mass. The top panel of \Fig{fig:AM_moster} shows the $\Mstar-\Mhalo$ relation for the entire suite of galaxies with the abundance matching prediction from \citet{Moster13} indicated as the central solid black line with the $1\sigma$ uncertainties plotted as thin lines above and below the central relationship. 
Each galaxy is colored according to the feedback model and symbol coded correspondingly to which initial condition was used, as described in Table 1.

Simulations are scattered around the $M_\star-\Mhalo$ relation. The fiducial feedback (red) represents the best fit to the abundance matching relation at every halo mass. Increasing the early stellar feedback efficiency $\epsilon_{\rm esf}$ (blue) reduces the stellar mass by a factor of two at the high mass end, while leaving the total amount of stars relatively unchanged at the low mass end, compared to the fiducial feedback. 
When early stellar feedback is not included the energy per supernova must be increased to $E_{\rm SN}=1.2$ in order to lower the stellar mass to the \citet{Moster13} relation (cyan). We note that the star formation history using such feedback is quite different from the fiducial runs, with more star formation at high redshift \cite[see][for details]{stinson13}. The yellow simulations that include high $\epsilon_{\rm esf}$ have systematically lower stellar-to-halo mass ratios, and also have high late time star formation. Finally, the original MUGS feedback (black) systematically forms too many stars at each halo mass.

The bottom panel of \Fig{fig:AM_moster} shows $\alpha$ as a function of halo mass, where $\Mhalo$ comes
from the full hydrodynamical simulation\footnote{Using $\Mhalo$ taken from
the dark matter only run provides similar results, as the halo mass
amongst DM and SPH simulations changes by only a few percent.}. The solid black line shows the theoretical expectation of $\alpha$ as a function of halo mass for the DM only case, as in \citet{maccio08} assuming a WMAP3 cosmology; the thin solid lines represent the scatter in the $c$-$\Mhalo$ relation.

\begin{figure}
  \includegraphics[width=3.5in]{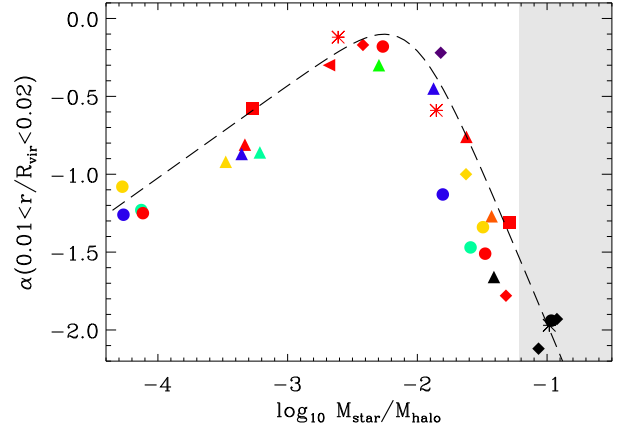}
 \caption{The relation between dark matter density profile slope, $\alpha$, measured between $0.01<r/\Rvir<0.02$, and the stellar-to-halo-mass ratio of each galaxy. Colors and symbols are the same as in \Fig{fig:AM_moster}. The best fit function of \Eq{fit_function} is overplotted as a dashed line. The grey area on the right side indicates the $1\sigma$ peak in the $\Mstar/\Mhalo$ abundance matching.}
\label{fig:ratio}
\end{figure}

At fixed halo mass, $\alpha$ varies greatly, depending on the feedback strength. The simulations that most closely follow the $\Mstar-\Mhalo$ relationship show a notable flattening of inner profile slopes as mass increases, as in \citet{Governato12}. This flattening is due to the increasing energy available from SNe explosions, as derived in \citet{Penarrubia12}. Indeed, all the galaxies in our sample whose inner slope is shallower than the corresponding DM run, have had an energy injection from SNe equal or higher than the conservative values found in \citet{Penarrubia12}. 
We note, however, that in our simulations the core creation process does not only depend on the total amount of energy available: in the g15784 MUGS dwarf galaxy (black triangle), for example, the energy from SNe is higher than in the g15784 dwarfs of the same mass that had an expansion, yet this galaxy is strongly contracted. What we observe is the interplay between the energy from stellar feedback and the increased potential well caused by the high number of stars at the galaxy center (see next section for more details).

The profiles are flattest around $\Mhalo\sim10^{11}\Msun$. 

At higher masses, however, the inner profiles steepen again. All the simulations above the $\Mstar-\Mhalo$ relationship have inner slopes $\alpha<-1.5$, i.e. a contracted halo steeper than the DM expectation at each halo mass. These simulations are all black colored indicating that they were part of the MUGS simulations.

Thus, depending on the feedback and the halo mass used, the dark matter haloes may expand, contract or retain the initial NFW inner slope. It seems that the inner slope of the dark matter density profile does not show a clear dependence on halo mass (or equivalently stellar mass) when different feedback schemes are included.

\subsection{Inner slope as a funcion of stellar-to-halo mass}

While there is not a well defined relation between $\alpha$ and stellar or halo mass individually, \Fig{fig:ratio} shows $\alpha$, measured between $0.01<r/\Rvir<0.02$, plotted as a function of $\Mstar/\Mhalo$. The dark matter inner profile slope shows a tight relationship as a function of $M_\star / \Mhalo$: indeed, much of the scatter apparent when $\alpha$ was plotted as a function of $\Mhalo$ disappears. The grey area indicates the region where the $M_\star / \Mhalo$ ratios are more than $1\sigma$ above the $\Mstar/\Mhalo$ peak in the abundance matching relation. Real galaxies do not have these star formation efficiencies.

The tight relationship between $\alpha$ and $M_\star / \Mhalo$ points to the conditions in which stellar feedback can create dark matter density cores. At low values of $\Mstar/\Mhalo$, the stellar content per halo mass is too small for the feedback energy to modify the DM distribution, and the halo of such galaxies retains a cuspy profile. As the stellar content per halo mass increases, the feedback energy is strong enough to produce expanded dark matter haloes, and thus for increasing values of $\Mstar/\Mhalo$ the inner slope of dark matter profiles gets flatter, reaching a maximum of $\alpha=-0.10$ at $\Mstar/\Mhalo=0.5$ per cent. The maximum value of $\alpha$ is even smaller, i.e. the profiles are flatter, if the inner slope is measured closer to the centre.  At $3<r/\epsilon<10$, $\alpha\sim0$ at $\Mstar/\Mhalo=0.35$ per cent. At higher masses, the amount of stars formed in the central regions deepens the potential well at the center of the galaxies, opposing the expansion process and leading to increasingly cuspy profiles for higher values of $\Mstar/\Mhalo$.

We verified this claim by studying in detail the medium mass version of g15784 for different choices of feedback parameters. 
We found that the stellar mass within 1 kpc is a good indicator of the minimum of the potential in each galaxy and that, as expected, the cored most version of g15784 (green triangle) has the shallowest potential well. Looking at the evolution of this galaxy, we observe that its SFR decreases with time and correspondingly the $\Mstar/\Mhalo$ value within 1 kpc is fairly constant at every redshift, reaching only 0.1 at $z=0$; the fraction of gas vs stars at the center is always very high, making possible the core creation since there is enough gas per total mass (or stellar mass) to be efficient in flattening the profile. 

This process does not occur in the cuspy version g15784 fiducial (red triangle), which has a constant SFR after 11 Gyrs and its $\Mstar/\Mhalo$ ratio within 1 kpc increases up to 0.4 at $z=0$: the increasing amount of stars at the center causes the gas vs stars ratio to become very low, therefore the gas available for the outflows is not sufficient to be effective at flattening the profile because the potential well has been deepened by the stars.

We note that the total amount of gas in the inner 1kpc is similar in both the cored and the cuspy medium mass versions of g15784: it is not the absolute amount of gas which regulates the cusp/core transition, but its relative value compared to the total (or stellar) inner mass.
We conclude that stellar mass at the galaxy center and in particular the ratio $\Mstar/\Mhalo$ is the most important quantity at indicating the deepening of the gravitational potential which balances the energy released from SNe.

The relationship shown in \Fig{fig:ratio} can be analytically modelled. We use a four parameter, double power law function, whose best fit is shown in \Fig{fig:ratio} as a dashed black line:

\beq
\alpha(X)= n - {\rm log}_{10}\left[\left(\frac{X}{x_0}\right)^{-\beta}+\left(\frac{X}{x_0}\right)^\gamma\right],
\label{fit_function}
\eeq

\noindent where $X=\Mstar/\Mhalo$ while $\beta$ and $\gamma$ are the low and high star forming efficiency slopes.
The best fit parameters, summarized in Table 2, were obtained using a $\chi^2$ minimization fitting analysis. The same dependence, but with different normalization, is obtained for the various criteria used to define the inner radial range, also shown in Table 2.

\begin{table}
\label{tab:fit}
 \caption{Best fit parameters and relative errors for the $\alpha$ vs $\Mstar/\Mhalo$ relation. The reduced Chi-Square is also listed.}
\begin{center}
\begin{tabular}{lccccc}
\hline
\hline
radial range  & n & ${\rm log}_{10}x_0$ & $\beta$ &$\gamma$&$\chi^2_r$ \\
\hline
$0.01<r/\Rvir<0.02$  &           $0.132$ & $-2.051$ & $0.593$&   $1.99$& $1.16$  \\
&                                                  $\pm0.042$ & $\pm0.074$ & $\pm0.086$ &  $\pm0.32$  \\
    
 $1<r/\kpc<2$                                  & $0.168$ & $-2.142$ & $0.699$&   $1.56$& $1.29$  \\
  &                                                            $\pm0.031$ & $\pm0.133$ & $\pm0.213$ &  $\pm0.12$  \\

 $3<r/\epsilon<10$                              & $0.231$ & $-2.209$ & $0.494$&   $1.49$& $1.28$  \\
  &                                                            $\pm0.043$ & $\pm0.064$ & $\pm0.055$ &  $\pm0.55$  \\
 
\hline
\end{tabular}
\end{center}
\end{table}

\begin{figure}
  \includegraphics[width=3.4in]{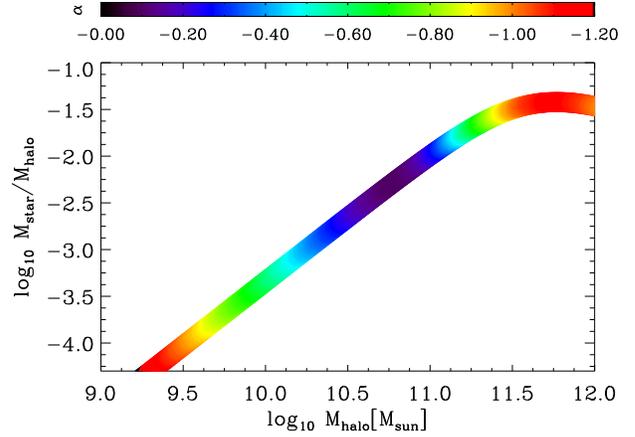}
  \caption{The abundance matching prediction color coded according to the expected value of the DM inner slope at every halo mass. We used the best fit parameters of $\alpha$ measured between $0.01$ and $0.02$ of each galaxy's virial radius.}
\label{fig:color}
\end{figure}	
	
\Fig{fig:color} shows the abundance matching relationship of $\Mstar/\Mhalo$ as a function of $\Mhalo$ color coded
according to the expected value of DM inner slope when $\alpha$ is measured at $0.01< r/\Rvir
< 0.02$.  The halo mass at which the flattest DM profiles are expected to be found, corresponding to a peak $\Mstar/\Mhalo=0.5$ per cent, is $\Mhalo\approx 10^{10.8}\Msun$. The profile becomes increasingly cuspy, approaching the NFW value for galaxies near the Milky Way mass: only galaxies with $\Mstar/\Mhalo>3.8$ per cent, which is the peak in the abundance matching prediction, are contracted. Such galaxies are outliers in the Universe.

\subsection{Core creation}\label{sec:creation}
We next examine which mechanism is responsible for the creation of cores, using the three simulations shown in \Fig{fig:profiles} as case studies.
As outlined in \S\ref{sec:introduction}, core formation from stellar feedback depends on repeated starbursts that are able to move gas enough to have a dynamical effect on the dark matter \citep{Read05,governato10,Maccio12,Pontzen12,teyssier13}. 
 
\begin{figure}
  \includegraphics[width=3.2in]{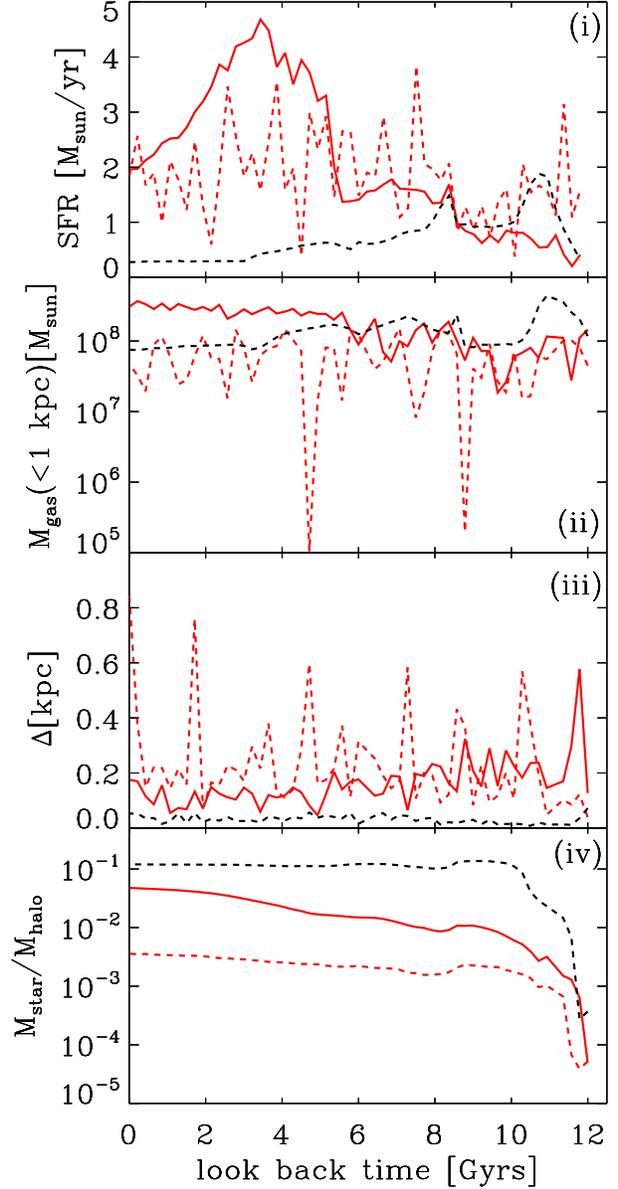}
  \caption{For the galaxies in Fig~\ref{fig:profiles}, we show the evolution of (i) the star
formation history; (ii) the gas flows within a 1 kpc sphere centered at the
galaxy center; (iii) the relative position between gas and dark matter potential
minima and (iv) the $\Mstar/\Mhalo$ as a function of lookback time. Note that the SFR of the $\Mstar=2.4\times 10^8\Msun$ galaxy (red dashed line) has been multiplied by a factor $100$ in order to be shown in the same scale range. }
\label{fig:comp}
\end{figure}

The four panels of \Fig{fig:comp} show how some relevant quantities vary as a function of lookback time. From top to bottom we present: (i) the star formation history, which shows clear starbursts that can drive outflows; (ii) the gas mass within a sphere of 1 kpc from the center of the galaxy, which shows when the gas has been driven out of the galaxy centre; (iii) the distance $\Delta$ between the position of the dark matter and gas potential minima, which shows how much the baryonic centre of mass moves around; and (iv) the $\Mstar/\Mhalo$ value that determines $\alpha$. 

The medium mass version of g5664 that uses the fiducial MaGICC feedback (red dashed line) has the flattest density profile at $z=0$, so we expect it to have the most violent history. Indeed, it has a bursty star formation history (multiplied by 100 to get it into the same range as the other galaxy star formation histories), and a star formation efficiency, $\Mstar/\Mhalo$, that stays near the optimal value for cores, between $\sim0.35$ and $0.5$ per cent throughout its evolution. A couple of the bursts of star formation cause significant gas loss from the inner 1 kpc, which results in consistent offsets between the positions of the center of gas and dark matter distributions. 

The medium mass version of g5664 that uses the low feedback MUGS physics (dashed black line) is the most contracted galaxy of this set. Other than a peak of star formation rate at an early time, corresponding to its peak dark matter accretion, its star formation history is a smoothly declining exponential. This early star formation quickly drives the efficiency $\Mstar/\Mhalo$ to values higher than $10$ per cent, which, according to \Fig{fig:ratio}, leads to a cuspy density profile. The high amount of stars already formed 11 Gyrs ago within this galaxy creates a deep potential well which suppresses the effects of stellar feedback, so that little gas flows out of the inner regions and the DM and gas distributions share the same centre of mass throughout the galaxy's evolution. 

Perhaps the most interesting case is that of the fiducial high mass g5664 galaxy (red solid line). At $z=0$ its dark matter profile is slightly contracted compared to the NFW halo, but less contracted than the lower mass MUGS case (dashed black line). Indeed, its star formation efficiency, $\Mstar/\Mhalo\sim5$ per cent at $z=0$, is lower than the MUGS case, but still high enough to have contracted dark matter. This galaxy shows elevated star formation starting $\sim$6 Gyrs ago, which correlates with an increase of $\Mstar/\Mhalo$, increased gas in the centre with fewer outflows and a more constant $\Delta$.
Before $z=0.66$ the star formation efficiency, $\Mstar/\Mhalo$, of this galaxy was still $\sim1$ per cent, and the feedback energy was still able to cause gas flows and variations in $\Delta$. When we examine the galaxy at that epoch, it indeed had an expanded dark matter profile with $\alpha> -1.0$, measured between $0.01$ and $0.02$ of the physical virial radius. Immediately after the starburst the star formation efficiency increases, the dark matter and gas start to share the same centre, the outflows from the inner region diminish, and the profile steepens to $\alpha<-1.0$ by $z=0.66$ (6 Gyrs ago) and finally to $\alpha=-1.8$ by $z=0$ with a star formation efficiency of $\Mstar/\Mhalo\sim5$ per cent.

\subsection{Predictions for observed galaxies}\label{sec:predictions}

\begin{figure}
  \includegraphics[width=3.4in]{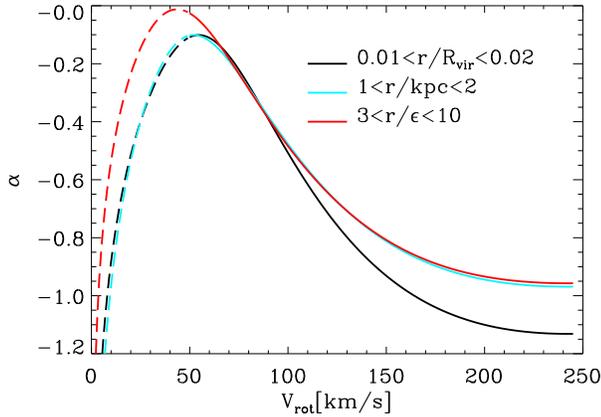}
  \caption{Expected relation between galaxies' rotation velocity and inner slope of their dark matter haloes. The three lines correspond to different radial ranges used for measuring $\alpha$. The dashed lines refer to the linear extrapolation of the baryonic TF relation \citep{Dutton10} below $\Mstar=10^9\Msun$.}
\label{fig:vrot}
\end{figure}

Combining the parameters in Table 2 with the \citet{Moster13} relationship, it is possible to predict the inner density profile slope of a galaxy based on its observed stellar mass.
This allows us to make predictions which are independent of the feedback
prescription. Using the best fit parameters from the $0.01<r/\Rvir<0.02$ range, we can compute the median expected $\alpha$ dependence on stellar mass for galaxies as massive as $\Mhalo\approx10^{12}\Msun$ ($\Mstar\approx 3.4\times 10^{10}\Msun$):

\begin{equation}
\alpha=0.132-\log_{10}\left[\frac {\eta^{2.58} +1}{\eta^{1.99}}\right]
\end{equation}
where
\begin{equation}
\eta = 0.84  \left({ {M_*} \over {10^9 {\rm M_{\odot}}}} \right)^{-0.58 }
+ 0.06  \left({ {M_*} \over {10^9 {\rm M_{\odot}}}} \right)^{0.26}
\end{equation}

\noindent The peak of this function occurs at $\Mstar=10^{8.5}\Msun$ and the low mass end slope, 0.34, is in good agreement with the one obtained in \citet{Governato12} for stellar masses between $10^4<\Mstar/\Msun<10^{9.4}$. Our study extends the prediction of cores vs cusps to L$^\star$ scales and predicts a turnover in the relation between inner slope and galaxy mass for $\Mstar>10^{8.5}\Msun$: above this value, the inner slope decreases as $\alpha \propto -0.64\log_{10}\Mstar/\Msun$.

Taking a step further, the stellar content of galaxies is then connected to their observed rotation velocity through the Tully-Fisher (TF) relation. 
Equation 4 of \citet{Dutton10} parameterizes \Vrot\ at 2.2 I-band exponential scale lengths as a function of $\Mstar$.
Using this $\Mstar-\Vrot$ relation we predict $\alpha$ as a function of $\Vrot$, the rotation velocity of galaxies. 
\Fig{fig:vrot} shows, for the different radial ranges where we measure the inner density profile, $\alpha$ as a function of observed rotation velocity for galaxies with $\Mhalo\leqslant
10^{12}\Msun$. The dashed lines indicate where the Tully-Fisher relationship was linearly extrapolated for $\Mstar<10^9\Msun$.

\Fig{fig:vrot} shows that the galaxies with the flattest inner density profiles are found at $\Vrot\sim50\,\kms$. $\alpha$ decreases in more massive galaxies where the inner density profiles become more cuspy until they reach the NFW profile. 

We note that the position at which the inner slope is measured has an effect on the $\alpha$ values, which alters the best fit parameters reported in Table 2, and consequently determines how $\alpha$ varies with rotation velocity.  Thus, \Fig{fig:vrot} has to be interpreted according to the radial range chosen, though the general trends are not changed and the peak of $\alpha$ remains at $\Vrot\sim50\,\kms$, independent of where the slope is measured.

The major difference between $\alpha$ measured at $0.01<r/\Rvir<0.02$ and the other radial ranges is that the inner slope is steeper for $\Vrot>100\,\kms$ in the former case.  A steeper slope is expected because $0.01<r/\Rvir<0.02$ is further from the galaxy centre than the other two measurements. However, none of the measured $\alpha$ values fall  below the NFW expectation as \Vrot\ approaches 250 \kms. Thus, dark matter haloes are never contracted in our model, even in the most massive disc galaxies.

\section{Conclusions}\label{sec:conclusion}
Using 31 simulated galaxies from the MaGICC project, we
showed that dark matter density profiles 
are modified by baryonic processes in the centre of galactic haloes.
The inner profile slope depends solely on the mass of stars formed per halo
mass and is
independent of the particular choice of feedback parameters within our blastwave and early stellar feedback scheme. Similar to previous work, the expansion of the dark matter profile results from supernova-driven outflows that cause fluctuations in the global potential and shift the centre of the gas mass away from the centre of the dark matter mass.

At  values of $\Mstar/\Mhalo\lsim 0.01$ per cent, the energy from stellar feedback is not sufficient to modify the DM distribution, and these galaxies retain a cuspy profile. At higher ratios of stellar-to-halo mass, feedback drives the expansion of the DM haloes, resulting in cored profiles. 
The shallowest profiles form in galaxies with $\Mstar/\Mhalo\sim 0.5$ per cent.  According to the abundance matching relation \citep{Moster13}, these galaxies have $\Mhalo \approx 10^{10.8}\Msun$ and $\Mstar\approx10^{8.5}\Msun$. 
In higher mass haloes, the deepening of the potential due to stars that form in the central regions suppresses supernova-driven outflows and thus lowers expansion, leaving cuspier profiles.

The abundance matching peak of star formation efficiency,
$\Mstar/\Mhalo=3.8$ per cent, occurs at $\Mhalo=10^{11.76}\Msun$, which is close to the
lowest current estimate of the Milky Way mass. Our model predicts that such a halo will be uncontracted and have an NFW-like inner slope of $\alpha=-1.20$ when the slope is measured between $\sim2$ and $\sim4\kpc$.

We combine our parameterization of $\alpha$ as a function of $\Mstar/\Mhalo$ with the empirical abundance matching relation to assign a median relationship between $\alpha$ and $\Mstar$.  The inner slope of the dark matter density profile increases with stellar mass to a maximum (most cored profile) at  $\Mstar\approx10^{8.5}\Msun$, before decreasing toward cuspier profiles at higher stellar masses.
Below $\Mstar\approx10^{8.5}\Msun$ the DM inner slope increases with stellar mass as $\alpha\propto 0.34\log_{10} \Mstar/\Msun$, similar to the relation found in \citet{Governato12}. For $\Mstar > 10^{8.5}\Msun$, dark matter haloes become cuspier, with $\alpha \propto -0.64\log_{10}\Mstar/\Msun$.

The Tully-Fisher relation allows us to predict the dependence of the DM inner slope on the observed rotation velocity of galaxies. Using our results and the stellar mass TF relation from \citet{Dutton10}, we find that the flattest inner profiles are expected for galaxies with $\Vrot \sim50\,\kms$. $\alpha$ decreases for more massive galaxies, leading to cuspier profiles and eventually reaching the NFW prediction at the Milky Way mass.
We note that, in agreement with our findings, the most clear observational measurements of flattened ``core" profiles of disc galaxies \citep{Deblok08,Kuzio08,Kuzio09,oh11b} are found in low surface brightness (LSB) galaxies with $\Vrot < 100\,\kms$. 

More massive disc galaxies, being baryon dominated, suffer from larger uncertainties in the disc-halo decomposition of their rotation curves, making it difficult to distinguish if their dark matter profile is cuspy or cored. Some studies conclude that such galaxies, those with $\Vrot > 150\,\kms$, can be described with cored profiles \citep{Borriello01,Donato04,McGaugh07}, while others find that NFW model provide equally good fits for these high luminosity galaxies \citep{Deblok08,Kuzio08}.

More recently, \citet{Martinsson13} presented rotation-curve mass decompositions of several massive spiral galaxies, and found no significant difference between the quality of a pseudo-isothermal sphere or a NFW model in fitting the DM rotation curves of individual galaxies, given the uncertainties in the contribution of baryons. 
If high surface brightness discs are sub-maximal \citep[e.g.][]{Courteau99} their haloes are allowed to be cuspy at the center. 

An aspect not taken into account in our simulations of galaxy formation is the influence of AGN feedback on the density profile of dark matter haloes. We acknowledge that this form of feedback starts to be relevant at the high halo mass end, where we observe increasingly cuspy galaxies: the study of the core/cusp problem would thus benefit from a future implementation of this type of feedback.

Our novel prediction for cusp vs core formation can be tested and, at least at the low halo mass end, well constrained using observational data sets.
This study can be applied to theoretical modeling of galaxy mass profiles, as well as to modeling of populations of disc galaxies within cold dark matter haloes. We find this encouraging, and hope that our study motivates more systematic analysis of the dependance of $\alpha$ on galaxy mass in real disc galaxies.

\section*{Acknowledgements}

The authors thank the referee for thoughtful comments on the manuscript. They further thank Julio Navarro, Fabio Governato, Andrew Pontzen, Paolo Salucci and Erwin de Blok for useful and constructive discussions.

ADC thanks the MICINN (Spain) for the financial support through the grant AYA2009-13875-C03-02 and the MINECO grant AYA2012-31101.
She further thanks the MultiDark project, grant CSD2009-00064.
ADC and CBB thank the Max- Planck-Institut f\"{u}r Astronomie (MPIA) for its hospitality.
CBB is supported by the MICINN through the grant AYA2009-12792. 
CBB, AVM, GSS, and AAD acknowledge support from 
the  Sonderforschungsbereich SFB 881 ``The Milky Way System'' 
(subproject A1) of the German Research
Foundation (DFG).  
AK is supported by the MICINN through the Ram\'{o}n y Cajal programme as well as the grants AYA 2009-13875-C03-02, AYA2009-12792-C03-03, CSD2009-00064, CAM S2009/ESP-1496 (from the ASTROMADRID network) and the MINECO through grant AYA2012-31101. He further thanks Ennio Morricone for adonai.
We acknowledge the computational support provided by the UK's National
Cosmology Supercomputer (COSMOS), the {\sc theo} cluster of the
Max-Planck-Institut f\"{u}r Astronomie at the Rechenzentrum in
Garching and the University of Central Lancashire's High Performance
Computing Facility.  
We thank the DEISA consortium, co-funded through EU FP6 project
RI-031513 and the FP7 project RI-222919, for support within the DEISA
Extreme Computing Initiative.


\bibliographystyle{mn2e}
\bibliography{archive}

\bsp

\label{lastpage}

\end{document}